# MEASUREMENTS OF ANISOTROPY IN THE COSMIC MICROWAVE BACKGROUND RADIATION AT DEGREE ANGULAR SCALES NEAR THE STARS SIGMA HERCULES AND IOTA DRACONIS

A. C. Clapp[1], M. J. Devlin[1], J. O. Gundersen[2], C. A. Hagmann[1], V. V. Hristov[1], A. E. Lange[1], M. Lim[2], P. M. Lubin[2], P. D. Mauskopf[1], P. R. Meinhold[2], P. L. Richards[1], G. F. Smoot[3], S. T. Tanaka[1] P. T. Timbie[1,4], and C. A. Wuensche[2,5]

[1]Department of Physics, University of California at Berkeley, Berkeley, CA 94720; also NSF Center for Particle Astrophysics.

[2]Department of Physics, University of California at Santa Barbara, Santa Barbara, CA 93106; also NSF Center for Particle Astrophysics.

[3]Physics Division, Lawrence Berkeley Laboratory, Berkeley, CA 94720; also NSF Center for Particle Astrophysics.

[4]Department of Physics, Box 1843, Brown University, Providence, RI 02912

[5]Instituto Nacional de Pesquisas Espaciais-INPE/MCT, Departmento de Astrofisica, Sao Jose dos Campos, SP, Brazil 12200

ABSTRACT

We present results from two four-frequency observations centered near the stars Sigma Hercules and Iota Draconis during the fourth flight of the Millimeter-wave Anisotropy eXperiment (MAX). The observations were made of 6° x 0°.6 strips of the sky with a 1°.4 peak to peak sinusoidal chop in all bands. The FWHM beam sizes were 0°.55±0°.05 at 3.5 cm$^{-1}$ and a 0°.75±0°.05 at 6, 9, and 14 cm$^{-1}$. Significant correlated structures were observed at 3.5, 6 and 9 cm$^{-1}$. The spectra of these signals are inconsistent with thermal emission from known interstellar dust populations. The extrapolated amplitudes of synchrotron and free-free emission are too small to account for the amplitude of the observed structures. If the observed structures are attributed to CMB anisotropy with a Gaussian autocorrelation function and a coherence angle of 25', then the most probable values are $\Delta T/T_{CMB} = 3.1^{+1.7}_{-1.3} \times 10^{-5}$ for the Sigma Hercules scan, and $\Delta T/T_{CMB} = 3.3^{+1.1}_{-1.1} \times 10^{-5}$ for the Iota Draconis scan (95% confidence upper and lower limits).

*Subject headings:* Cosmic microwave background - cosmology: observations

## 1. INTRODUCTION

Measurements of the anisotropy of the cosmic microwave background (CMB) provide an effective method for testing and constraining models of cosmic structure formation. The Cosmic Background Explorer (COBE) satellite has detected anisotropy at large angular scales (Smoot et al. 1992). Recently, there has been a concerted effort to measure medium scale anisotropy (Schuster et al. 1993; Gaier et al. 1992; Wollack et al. 1993, Cheng et al. 1993). The results of CMB observations during the second, third, and fourth balloon flights of the Millimeter-wave Anisotropy eXperiment (MAX) were reported in Alsop et al. (1991), Gundersen et al. (1993), Meinhold et al. (1993a), and Devlin et al. (1994). We report here on two medium scale CMB anisotropy observations made during the fourth flight of MAX in low dust emission regions near the stars Sigma Hercules and Iota Draconis.

## 2. MEASUREMENT

The instrument has been described in detail elsewhere (Fischer et al. 1992; Alsop et al. 1992; Meinhold et al. 1993b). It consists of an off-axis Gregorian telescope and a bolometric photometer mounted on an attitude-controlled balloon platform which makes measurements at an altitude of 36 km. A new single pixel four band bolometric receiver was used for this flight. It features greatly reduced sensitivity to radio frequency (RF) interference, an additional frequency band at 3.5 $cm^{-1}$, and an adiabatic demagnetization refrigerator to cool the bolometric photometer to 85 mK (Clapp et al. 1993). The underfilled optics provide a 0°.55±0°.05 FWHM beam in the single mode 3.5 $cm^{-1}$ band and 0°.75±0°.05 FWHM beams in the multi-mode 6, 9, and 14 $cm^{-1}$ bands. The bandwidths are $\delta\nu/\nu$ = 0.57, 0.45, 0.35 and 0.25 FWHM, respectively. In order to convert antenna temperatures measured in the 3.5, 6, 9, and 14 $cm^{-1}$ bands to 2.726 K thermodynamic temperatures, multiply the antenna temperatures by 1.54, 2.47, 6.18, and 34.08, respectively.

The observations consisted of constant velocity scans in azimuth of ±3°.0 on the sky while tracking the pointing stars Sigma Hercules $\alpha = 16^h30^m$, $\delta = 42°46'$ and Iota Draconis $\alpha = 15^h25^m$, $\delta = 59°36'$ (Epoch 1993). A complete scan from 3° to -3° and back to 3° required 108 seconds. These regions were chosen because the IRAS 100 μm map (Wheelock S. et al. 1991) indicates they have low dust emission. The Sigma Hercules scan lasted from UT = 8.73 to UT = 9.44 hours. The Iota Draconis scan lasted from UT = 7.15 to UT = 7.60 hours, both on June 15, 1993. Calibrations were made before and after each observation using the membrane transfer standard described in Fischer et al. (1992). Scans of Jupiter were made between UT = 5.08 and 5.33 hours to confirm the calibration. The Jupiter calibration agrees with the membrane calibration to 10% in %%each band. Hence we assume a 10% error in the absolute calibration. The instrument is %%calibrated so that a chopped beam centered at the boundary between regions with temperatures $T_1$ and $T_2$, would yield $\Delta T = T_1 - T_2$.

### 3. DATA REDUCTION AND ANALYSIS

Transients due to cosmic rays were removed using an algorithm described in Alsop et al. (1992), which removed 15 - 20% of the data. The detector %%output was demodulated using the sinusoidal reference from the nutating secondary to produce antenna temperature differences, $\Delta T_A$, on the sky. The phase synchronous demodulation produced a measurement for each nutation cycle (every 185 ms) except when a cosmic ray was detected. No RF interference was observed. The noise averaged over an observation gives effective sensitivities of 580, 545, 770, and 2660 μK√sec in CMB thermodynamic units in the 3.5, 6, 9, and 14 $cm^{-1}$ bands, respectively. Due to noise from a faulty JFET, only 30% of the 6$cm^{-1}$ data from the Sigma Hercules scan, and 37% of the 6$cm^{-1}$ data from the Iota Draconis scan, were useable.

Each of the bands has an instrumental offset, originating from emissivity differences on the primary mirror and from atmospheric emission. The average of the measured offsets in antenna temperature were 3.0, 1.2, 0.9, 0.7 mK at 3.5, 6, 9, and 14 cm$^{-1}$. In some of the bands the value of the offset varied smoothly as a function of time. No offset drift was observed in the 3.5 cm$^{-1}$ band. The offset and offset drifts in each band were removed by a linear least squares fit to each half scan. This procedure was valid because the 54 second time scale of the fit was smaller than that of the offset drifts. The reduction in the rms amplitude of the signals in the 6, 9, and 14 cm$^{-1}$ bands was consistent with the reduction in the noise. The noise and rms amplitude in the 3.5 cm$^{-1}$ band were uneffected. This procedure removes any offset or gradient from the astrophysical signal.

For each scan, the means, variances, and 1$\sigma$ error bars of the antenna temperature differences were calculated for 21 pixels separated by 17' on the sky. The data are available from the authors. Most of the essential features of the data are apparent in Figures 1A and 1B. In both observations there is statistically significant correlated structure in the 3.5, 6, and 9 cm$^{-1}$ bands. The amplitude of the structures, as measured in antenna temperature, decrease with increasing frequency. There is no significant signal at 14 cm$^{-1}$ in the Sigma Hercules scan, but there is a small uncorrelated signal at 14 cm$^{-1}$ in the Iota Draconis scan. The correlations for the Sigma Hercules scan are 0.55 (3.5-6), 0.46 (3.5-9), 0.21 (6-9). The correlations for the Iota Draconis scan are 0.33 (3.5-6), 0.43 (3.5-9), 0.42 (6-9), -0.03 (3.5-14), 0.12 (6-14), 0.20 (9-14). Both the values of the rms $\Delta T_A$, corrected for the contribution from detector noise, and the probability of reproducing the measured rms $\Delta T_A$ with gaussian random noise are shown in Table 1. The error on each rms includes a statistical error and a $\pm$10% estimate for the uncertainty in the calibration.

In order to further test the hypothesis that the signals in the 3.5, 6, and 9 cm$^{-1}$ bands are correlated, a best fit model was determined by minimizing

$$\chi_R^2 = \sum_{j=1}^{3} \sum_{i=1}^{21} (x_{ij} - a_j y_i)^2 / \sigma_{ij}^2. \qquad (1)$$

Here $x_{ij}$ and $\sigma_{ij}$ are the measured means and variances of the 21 pixels, $y_i$ represents the best-fit sky model, and $a_j$ the best fit model scale factor for band j. For Sigma Hercules, the probability of exceeding the residual $\chi_R^2$ is 0.20. For Iota Draconis, the probability of exceeding the residual $\chi_R^2$ is 0.48. These good three-band fits indicate that essentially all of the signal in excess of noise in these bands is correlated and consistent with a single source of emission. The spectral analysis of the signals must take into account the 0°.55 FWHM beam in the 3.5 cm$^{-1}$ band compared with the 0°.75 FWHM beam in the other bands. Table 2 shows the ratio of the signal amplitudes in our bands, and compares them with the ratios predicted by several relevant sky models which specify both the spectrum and angular distribution of emission.

## 4. POTENTIAL SYSTEMATIC ERRORS

All anisotropy experiments are potentially susceptible to off-axis response from a variety of sources including the Sun, the Moon, the Earth, the balloon, and the Galactic plane. During the entire MAX4 flight, the Sun and Moon were both below the horizon. The unchopped off-axis response of this instrument has been measured from 0° to 35° in elevation below the boresight. The unchopped off-axis response is ≥ 65 dB below the on-axis response at angles from 13° to 35°. Baffles have been used in each flight of the MAX experiment to minimize the off-axis response of the instrument. The design of the baffles was modified for this flight to ensure that no direct or reflected emission from the Earth or balloon could directly illuminate the optical system.

During this flight of the MAX experiment, a region near the star Gamma Ursa Minoris (GUM) was also observed (Devlin et al., 1994). This region of the sky has been observed three times by MAX, each with substantially altered baffles, and in this flight with a new receiver and different beam sizes. The elevation of GUM ranged from near 30° in the first observation, near 35° in the second observation to near 48° in the third observation. In each case the same amplitude signal was observed, with a spectrum consistent with CMB. It is unlikely that sidelobe contamination would be reproducible at this level under such drastically changed observing conditions.

## 5. FOREGROUND SOURCES

Potential foreground sources of confusion include the atmosphere, interstellar dust emission, synchrotron radiation, free-free emission, the Sunyaev-Zel'dovich (SZ) effect, and radio point sources. The amplitude of the emission in the 14 cm$^{-1}$ band provides stringent limits on flat or rising spectra such as emission from 20 K galactic dust, ambient temperature objects, or the atmosphere. Based on an extrapolation from the IRAS 100 μm data (Wheelock et al. 1991) and scaling the brightness by the spectrum for high latitude dust reported by Meinhold et al. (1993a), the differential dust emission in the Sigma Hercules and Iota Draconis regions is expected to be a factor of 2 below the measured structures at 14 cm$^{-1}$ and a factor of 60-100 below the measured structures at 3.5 cm$^{-1}$. The observed anisotropy is inconsistent with the thermal SZ effect, which would produce anti-correlated structure at 6 and 9 cm$^{-1}$.

The rms differential antenna temperature due to diffuse synchrotron emission at 408 MHz in the Sigma Hercules region is <0.77K, as determined by convolving our scan pattern with the 30' × 30' smoothed version of the Haslam et al. (1982) map. Assuming a scaling law $\Delta T_A \propto \nu^\beta$ for synchrotron emission with $\beta$ = -2.7 implies an rms $\Delta T_A \leq$ 0.38, 0.06, 0.02, and 0.006 μK at 3.5, 6, 9, and 14 cm$^{-1}$, respectively. These estimates account for <1% of the observed structure, with similar results for the %%Iota Draconis scan.

A very conservative estimate of the amplitude of free-free emission is obtained from the assumption that the entire 408 MHz rms is due to free-free emission and extrapolating to our frequencies using $\Delta T_A \propto \nu^{-2.1}$. This estimate gives 9.2 μK rms at 3.5 cm$^{-1}$ for the Sigma Hercules scan, or <15% of the measured rms, with similar %%results (<25% of the measured rms) for the Iota Draconis scan. These amplitude considerations show that free-free emission is not likely to be a large contaminant of the observed signals. A thorough catalogue search has been made for bright or inverted spectrum radio sources in these regions. There are no radio point sources of sufficient brightness to produce a measurable signal in either of these scans.

## 6. DISCUSSION

Since all of the known possible foreground contaminants are considered unlikely, and the spectrum of the structure is consistent with CMB anisotropy, the following discussion interprets the observed structure as CMB anisotropy. The rms of the data give $\Delta T_{rms}/T_{CMB} = 3.2 \pm 0.7 \times 10^{-5}$, $1.20 \pm 0.8 \times 10^{-5}$ and $2.5 \pm 0.8 \times 10^{-5}$ (68% confidence level) for the Sigma Hercules scan, and $\Delta T_{rms}/T_{CMB} = 2.3 \pm 0.8 \times 10^{-5}$, $1.7 \pm 0.8 \times 10^{-5}$ and $2.1 \pm 0.8 \times 10^{-5}$ (68% confidence level) for the Iota Draconis scan, in the 3.5, 6, and 9 cm$^{-1}$ bands respectively. By integrating the window function for each of the bands over a Cold Dark Matter (CDM) model with $\Omega_B = 0.03$ and h=0.5 (Sugiyama and Gouda, 1993) normalized to the rms amplitude of the anisotropy measured by COBE at $\theta > 10°$ (Smoot et al. 1992) a prediction of the rms amplitudes expected for MAX can be made. For the 3.5 cm$^{-1}$ band this is $\Delta T_{rms}/T_{CMB} = 2 \times 10^{-5}$ and for the 6 and 9 cm$^{-1}$ bands $\Delta T_{rms}/T_{CMB} = 1.6 \times 10^{-5}$. The sampling variance, due to the small fraction of sky covered, expected for a single MAX scan is ≈25% (Scott et al. 1994).

We have analyzed the data assuming the CMB anisotropy is described by a Gaussian autocorrelation function (GACF) with a coherence angle of 25'. For measurements near the Doppler peak of a standard recombination CDM model, this is a useful approximation. A description of the Gaussian autocorrelation function analysis used for these data is given in Cheng et al. (1994). The GACF analysis gives most probable values in the 3.5, 6, and 9 cm$^{-1}$ bands of $\Delta T/T_{CMB} = 3.8^{+2.7}_{-1.9} \times 10^{-5}$, $1.2^{+3.2}_{-1.0} \times 10^{-5}$, and $2.6^{+2.8}_{-1.7} \times 10^{-5}$ for the Sigma Hercules scan, and $\Delta T/T_{CMB} = 3.4^{+1.1}_{-1.8} \times 10^{-5}$, $3.3^{+3.5}_{-2.1} \times 10^{-5}$, and $2.8^{+3.4}_{-2.0} \times 10^{-5}$ for the Iota Draconis scan (95% confidence upper and lower %%limits), respectively. These three numbers can be combined in a simple way by multiplying a fit to the likelihood functions to give $\Delta T/T_{CMB} = 3.1^{+1.7}_{-1.3} \times 10^{-5}$ for the Sigma Hercules scan and $\Delta T/T_{CMB} = 3.3^{+1.1}_{-1.1} \times 10^{-5}$ for the Iota Draconis scan (95% confidence upper and lower limits).

A complete summary of all medium scale CMB anisotropy results from the MAX experiment is given in Table 3. All these values were determined assuming a CMB described by a GACF with a coherence angle of 25'. All of the scans are statistically consistent, except for the MAX3 Mu Pegasus scan. This scan was different from the others, in that a significant galactic dust signal was present in the data, and was subtracted in the analysis. Calculating a weighted mean and uncertainty in that mean for the six MAX measurements in Table 3, the result is $\Delta T/T_{CMB} = 2.9 \pm 0.5 \times 10^{-5}$. The $\chi^2$ for these six measurements is then 16.9, with a probability of exceeding this $\chi^2$ of 0.5%. This low probability could indicate either a systematic error in the experiment, or that the assumed model of a guassian sky is incorrect. If the MAX3 Mu Pegasus scan is excluded, the weighted mean and uncertainty in that mean is $\Delta T/T_{CMB} = 3.6 \pm 0.2 \times 10^{-5}$. The $\chi^2$ for these five measurements is 2.5, with a probability of exceeding this $\chi^2$ of 63%.

## 7. CONCLUSION

We have presented new results from a search for CMB anisotropy with high sensitivity at angular scales near 1 degree. Significant structure is detected in the 3.5, 6 and 9 cm$^{-1}$ bands. If all of the structure is attributed to CMB anisotropy with a Gaussian autocorrelation function and coherence angle of 25', then the most probable values are $\Delta T/T_{CMB} = 3.1^{+1.7}_{-1.3} \times 10^{-5}$ for the Sigma Hercules scan, and $\Delta T/T_{CMB} = 3.3^{+1.1}_{-1.1} \times 10^{-5}$ for the Iota Draconis scan respectively (95% confidence upper and lower %%limits). Based on spectral and temporal arguments, sidelobe contamination from the Earth, balloon and the Galaxy are considered unlikely to cause the observed structure. The data rule out Galactic dust emission via the spectrum, morphology and amplitude of the structure. Synchrotron and free-free emission are considered unlikely contaminants from estimates of their intensity based on low-frequency maps. The spectrum of the signals at 3.5, 6, 9 and 14 cm$^{-1}$ is consistent with CMB anisotropy.

This work was supported by the National Science Foundation through the Center for Particle Astrophysics (cooperative agreement AST-9120005), the National Aeronautics and Space Administration under grants NAGW-1062 and FD-NAGW-2121, the University of California, and previously California Space Institute. Thanks to Prof. E. Haller for the NTD thermistors used for the bolometers.

TABLE 1
Signal RMS in Antenna Temperature, and
Probability of The Observed Signals

| Sigma Hercules Data | | |
|---|---|---|
| Band | Signal RMS | Probability |
| 3.5 cm$^{-1}$ | 57 ± 14 μK | <10$^{-9}$ |
| 6 cm$^{-1}$ | 13 ± 8 μK | 0.30 |
| 9 cm$^{-1}$ | 11 ± 3 μK | 0.05 |
| 14 cm$^{-1}$ | 0 ± 2 μK | 0.83 |
| Iota Draconis Data | | |
| 3.5 cm$^{-1}$ | 41 ± 14 μK | <10$^{-3}$ |
| 6 cm$^{-1}$ | 19 ± 9 μK | 0.20 |
| 9 cm$^{-1}$ | 9 ± 4 μK | 0.06 |
| 14 cm$^{-1}$ | 4 ± 2 μK | 0.24 |

TABLE 2
Ratios of the best fit signal amplitudes for pairs of bands compared with theoretical models of emission.

| Channel Ratio | 6/3.5 | 9/3.5 |
|---|---|---|
| Best-Fit Antenna Temp.[a] for SIGMA HERC. | 0.34 ±0.12 | 0.12 ±0.06 |
| Best-Fit Antenna Temp.[a] for IOTA DRACONIS | 0.46 ±0.16 | 0.20 ±0.07 |
| CMB[b] | 0.62 | 0.25 |
| CMB CDM[c] | 0.50 | 0.20 |
| CMB Unresolved Source[d] | 0.34 | 0.13 |
| Free-Free[b] | 0.37 | 0.16 |
| Free-Free Unresolved Source[d] | 0.20 | 0.08 |

[a] Best-fit ratio ± 1 $\sigma$ error. This includes a 10% calibration error.
[b] Ratios calculated with no beam size correction.
[c] Ratios calculated using Cold Dark Matter power spectrum (Sugiyama et al. $\Omega_B = 0.03$, h=0.5) to correct for beam size difference.
[d] Ratios calculated for a point source response. These are considered a lower limit. The uncorrected ratios[b] are upper limits.

TABLE 3
Summary of MAX results: Most probable values of measured signal amplitudes $\Delta T/T_{CMB}$, assuming a Gaussian auto-correlated sky with a coherence angle of 25' (95% confidence upper and lower limits).

| | |
|---|---|
| MAX2 GUM | $4.5^{+5.7}_{-2.6} \times 10^{-5}$ |
| MAX3 µPeg | $1.5^{+1.1}_{-0.7} \times 10^{-5}$ |
| MAX3 GUM | $4.2^{+1.7}_{-1.1} \times 10^{-5}$ |
| MAX4 GUM | $3.7^{+1.9}_{-1.1} \times 10^{-5}$ |
| MAX4 Iota Draconis | $3.3^{+1.1}_{-1.1} \times 10^{-5}$ |
| MAX4 Sigma Hercules | $3.1^{+1.7}_{-1.3} \times 10^{-5}$ |

FIGURE CAPTIONS

FIG. 1A,1B.--- Antenna temperature differences ($\pm$ 1 $\sigma$) for the 0.71 hours of data near Sigma Hercules (1A) and the 0.45 hours of data near Iota Draconis (1B). Each point is separated by 17' in azimuth. The solid lines show the response to a point source in each of the bands.

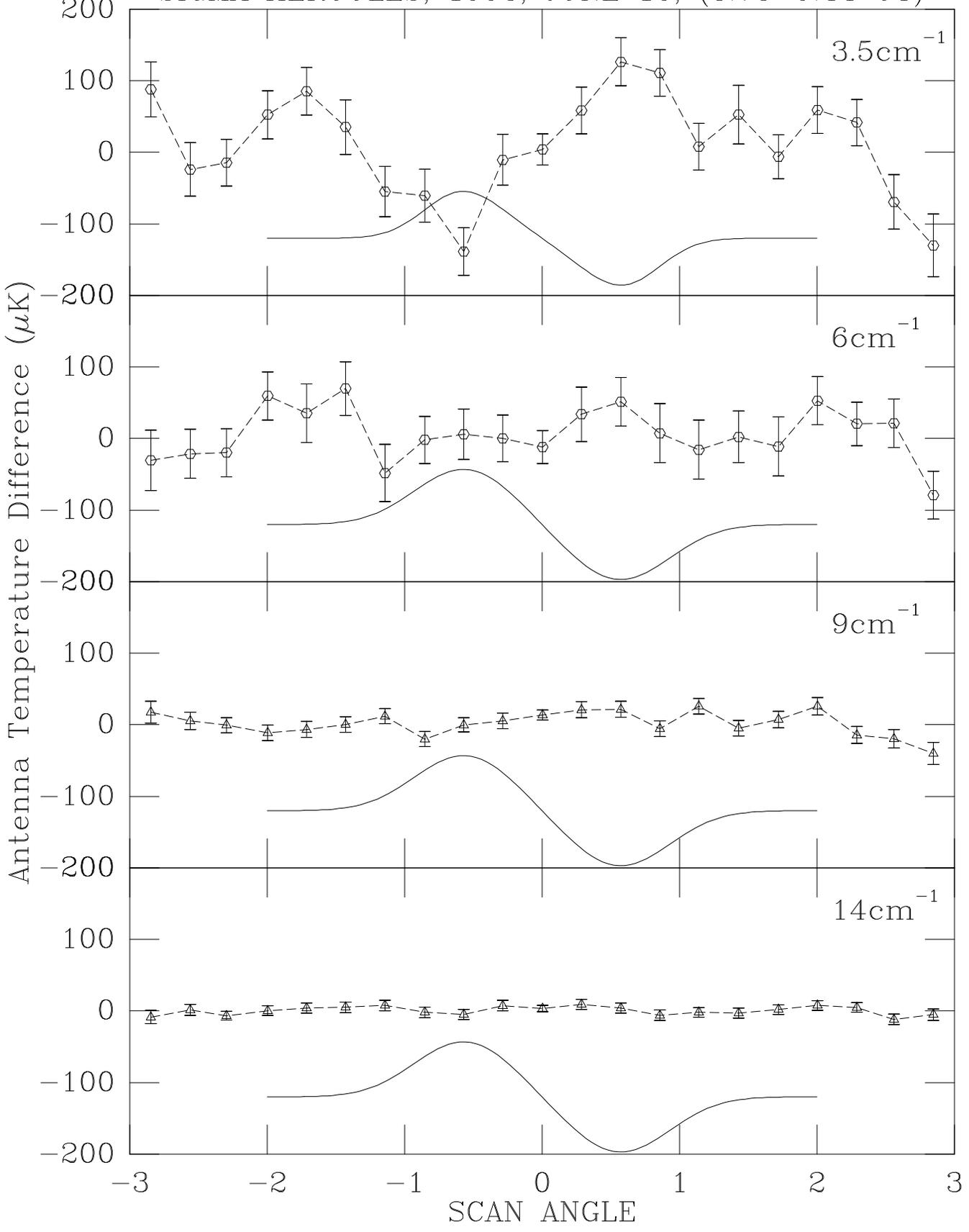

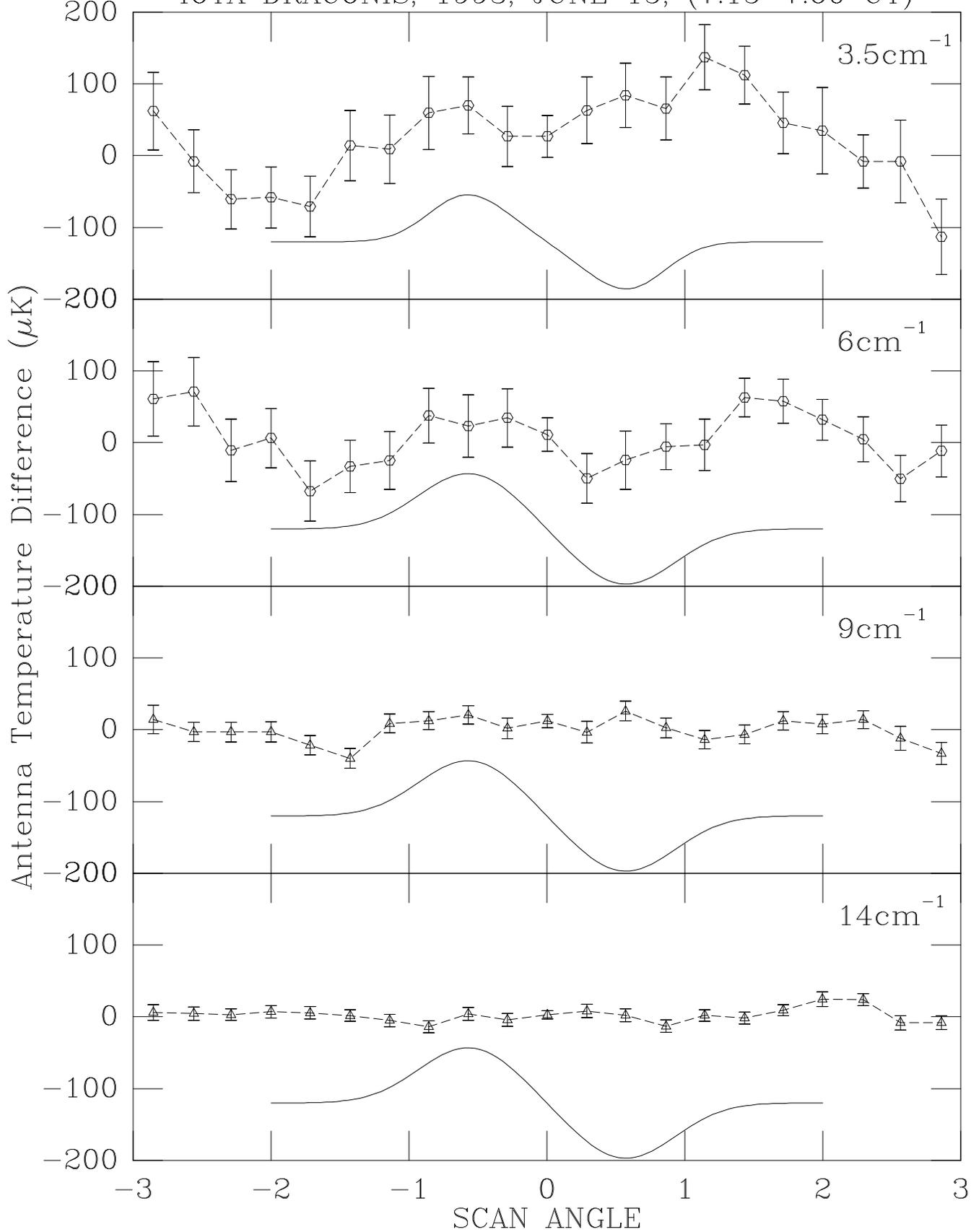